\documentclass[conference]{IEEEtran}

\usepackage[cmex10]{amsmath}
 
\DeclareMathOperator*{\argmin}{arg\,min} 

\usepackage{setspace}
\usepackage{graphicx}
\usepackage{subfigure}
\usepackage{tabularx}
\usepackage{arydshln}
\usepackage{umoline}
\usepackage{picinpar}
\usepackage{mathrsfs}
\usepackage{latexsym}
\usepackage{amssymb}
\usepackage{pifont}
\usepackage{amsfonts}
\usepackage{amsmath}
\usepackage{color}
\usepackage{fancyhdr}
\usepackage{float}
\usepackage{cases}
\usepackage{bm}
\usepackage{multirow}
\usepackage[english]{babel}
\usepackage{epsfig}
\usepackage{graphics}
\usepackage[Lenny]{fncychap}
\usepackage{rotfloat}
\usepackage{xcolor}
\usepackage{algorithm}
\usepackage{algpseudocode}

\newcommand\blfootnote[1]{%
  \begingroup
  \renewcommand\thefootnote{}\footnote{#1}%
  \addtocounter{footnote}{-1}%
  \endgroup
}

\usepackage{tabularx,booktabs}
\newcolumntype{C}{>{\centering\arraybackslash}X} 

\ifCLASSINFOpdf
\else
\fi

\newcommand{\ignore}[1]{}

\newcommand{\bigzeroU}{{\lower2ex\hbox{\huge 0}}}
\newcommand{\bigzeroL}{{\lower-.1ex\hbox{\huge 0}}}

\def\Vec#1{\boldsymbol{#1}}
\begin{document}
\title{Precoded Non-Orthogonal Frequency Division Multiplexing with Subcarrier Index Modulation}
\author{
\IEEEauthorblockN{Prakash~Chaki,~Takumi~Ishihara, and Shinya Sugiura$^*$}
\IEEEauthorblockA{Institute of Industrial Science, The University of Tokyo\\
Meguro-Ku, Tokyo 153-8505, Japan\\
Email: \{prakash.chaki@ieee.org, t.ishihara@ieee.org, sugiura@iis.u-tokyo.ac.jp\}}
}
\markboth{}
{Shell \MakeLowercase{\textit{et al.}}: Bare Demo of IEEEtran.cls for Journals}
\maketitle
\begin{abstract}
In this paper, we propose a novel precoded non-orthogonal frequency division multiplexing (NOFDM) with subcarrier index modulation (SIM) to increase spectral efficiency for multicarrier systems. The proposed NOFDM-SIM scheme is constituted by a filterbank multicarrier system under reduced subcarrier spacing, where precoded index modulated symbols are mapped onto the subcarriers. The detrimental effects of inter-carrier interference (ICI) resulting from the non-orthogonality between subcarriers are efficiently mitigated with the aid of eigenvalue-decomposition-based precoding. Our simulation results demonstrate that the proposed precoded NOFDM-SIM system exhibits a similar peak-to-average power ratio to OFDM and OFDM-SIM counterparts while achieving the bit error ratio limit of zero-ICI OFDM.
\end{abstract}

\IEEEpeerreviewmaketitle

\section{Introduction}
\label{intro}
\blfootnote{Preprint (accepted version) for publication in \textit{IEEE 95th Vehicular Technology Conference (IEEE VTC2022-Spring)}, 2022, DOI: 10.1109/VTC2022-Spring54318.2022.9860537. 
$\copyright$ 2022 IEEE. Personal use of this material is permitted. Permission from IEEE must be obtained for all other uses, in any current or future media, including reprinting/republishing this material for advertising or promotional purposes, creating new collective works, for resale or redistribution to servers or lists, or reuse of any copyrighted component of this work in other works.}
Orthogonal frequency-division multiplexing (OFDM)~\cite{hanzo_ofdm} is a multicarrier transmission technique, widely employed in diverse industry standards, such as digital audio broadcasting, digital video broadcast, asymmetric digital subscriber line, powerline communications, 4G/5G mobile communication, IEEE$802.11$ series.
A key advantage of OFDM driving its widespread use is its ability to convert a broadband dispersive channel into multiple narrowband frequency-flat subchannels.
However, the maximum spectral efficiency of OFDM is achievable when $TF{=}1$, where $T$ is symbol interval and $F$ is subcarrier spacing~\cite{gabor_book_chap}. Since Gabor's theory requires $TF{>}1$ for time-frequency localization, the spectral efficiency of OFDM is compromised. Furthermore, OFDM suffers from high out-of-band (OOB) emissions.
To combat the above-mentioned limitations of OFDM, several class of non-orthogonal frequency division multiplexing (NOFDM), including the filter bank multicarrier (FBMC)~\cite{fbmc-nissel}, generalized FDM (GFDM)~\cite{gfdm-michailow}, universal filtered multicarrier (UFMC)~\cite{ufmc-vakilian}, spectrally-efficient FDM (SEFDM)~\cite{sefdm-kanaras}, have been developed.
In FBMC, by filtering each subcarrier using narrowband orthogonal shaping filters, the spreading of side-lobes, i.e., OOB emission, is reduced.
Furthermore, owing to its lower sensitivity to carrier frequency offset relative to OFDM, the FBMC system typically performs better than OFDM in a high-mobility scenario.
UFMC offers an intermediate choice between OFDM and FBMC, where in UFMC, each clustered set of subcarriers is filtered, which allows a shorter filter length than in FBMC. 
In SEFDM, subcarrier spacing in the frequency domain is set lower than that of OFDM to improve spectral efficiency. This benefit is achieved at the cost of detrimental inter-channel interference (ICI) effects in order to combat this limitation, the zero-forcing precoding~\cite{xu2019jiot} and the eigenvalue decomposition (EVD)-precoded diagonalization~\cite{xu2019pimrc,osaki-svd-nofdm} were developed. Also, in~\cite{xu2019pimrc,osaki-svd-nofdm}, optimal power allocation employing the amplification factors formulated as the inverses of the eigenvalues was presented. Moreover, EVD-based SEFDM was extended to that supporting physical-layer security transmission~\cite{sugiura2021twc}.

Index modulation (IM)~\cite{sugiura_im} is a modulation technique that conveys information by activating subset indices of available slots in the space~\cite{sugiura2012universal,renzo-sm}, time~\cite{nakao_SCIM}, and frequency domains~\cite{ishikawa-sim}.
Specifically, a subset of available resource slots are activated for symbol transmission, while a part of the information bits is modulated using the subset activation pattern, and the rest is modulated onto conventional amplitude and phase-shift keying (APSK) symbol.
Additionally, IM supporting combinations of multiple resource domains, such as space-time domain \cite{sugiura-gstsk,kadir-space-time-freq-shiftkey} and space-time-frequency domains \cite{hoang2011spl}, have been developed. More recently, in \cite{nakao-sefdm-im}, the SEFDM with SIM (SEFDM-SIM) scheme was first studied without employing any precoding, and the minimum mean-square error (MMSE)-based linear receiver was developed. Also, joint channel estimation and equalization technique was developed for SEFDM-SIM~\cite{ma2020tcom}.
Furthermore, the time-domain counterpart of SEFDM-SIM was independently developed as faster-than-Nyquist (FTN)-IM~\cite{takumi-ftn-im}--\nocite{chaki2021isit}\cite{ishihara2021evolution}.
To the best of our knowledge, the conventional SEFDM-SIM scheme has not been implemented with any sophisticated precoding as well as power allocation.

Against the above background, the novel contribution of this paper is as follows.
We propose a novel precoded NOFDM-SIM scheme to enhance spectral efficiency exploiting the combined benefits of reduced subcarrier spacing of SEFDM while introducing the concept of SIM. More specifically, EVD-based precoding is employed to eliminate the effects of ICI imposed by the non-orthogonality of subcarriers. We present the bit-error ratio (BER) and peak-to-average power ratio (PAPR) performance of the proposed scheme for a channel-uncoded scenario. 

The rest of this paper is organized as follows. Section \ref{proposed-system-model} describes our proposed scheme, and Section \ref{performance_analysis} presents the performance analysis of the proposed scheme. 
Finally, Section \ref{sec:conclusion} concludes the present paper.
\section{System model of proposed scheme}
\label{proposed-system-model}

\begin{figure*}[!t]
\centering
\includegraphics[scale=0.7]{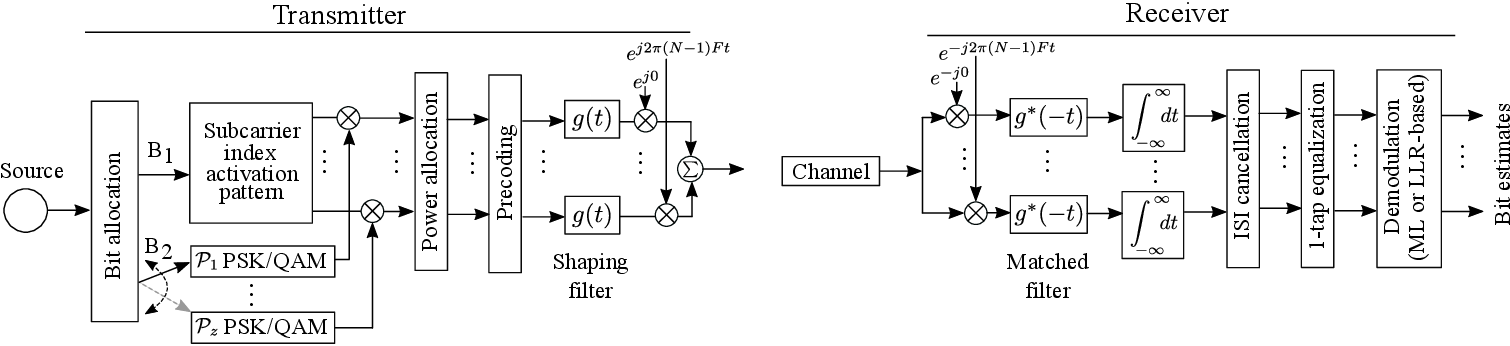}
\caption{Block diagram of proposed precoded NOFDM-SIM scheme.}
\label{fig:uncoded_block_dia}
\end{figure*}

Let us consider a precoded NOFDM-SIM scheme operating over an additive white Gaussian noise (AWGN) channel. The block diagram of our transceiver is shown in Fig. \ref{fig:uncoded_block_dia}. In the following subsections, we expound on the features of each block.

\subsection{Frequency-Domain Index Modulation}
\label{sec:system_model:modulation}
Information bits are modulated onto $N$ subcarriers, which are divided into $L$ clusters, each containing $M$ subcarriers. Thus, we have the relationship of $N=LM$.
As shown in Fig.~\ref{fig:uncoded_block_dia}, the subcarrier spacing is set to $F=\tau F_0$ $(0\leq\tau<1)$ where $F_0={1}/{T_0}$ and $T_0$ denote the subcarrier spacing and the symbol duration of OFDM, respectively.
We assume that $B$ information bits modulated onto each cluster, hence having a total of $LB$ information bits per frame.
For the $l$th cluster, $B$ information bits are divided into two groups, $B_1$ and $B_2$, where we have $B=B_1+B_2.$
The first $B_1$ bits are used to activate an index activation pattern of subcarriers, which corresponds to SIM. For example, if $K$ out of $M$ subcarriers are activated in a given cluster, then $B_1=\left\lfloor\log_2{M\choose K}\right\rfloor$ bits are modulated using based on the SIM principle.
Since $M-K$ subcarriers are deactivated, the symbols modulated on the active subcarriers are scaled by a factor $\sqrt{M/K}$ to maintain the same transmit power.
The remaining $B_2$ bits are modulated based on $\mathcal{P}$-ary APSK modulation, where $\mathcal{P}=2^{B_2}$.
Thus, the proposed modulation scheme is characterized by the parameter set of 
$(M,K,\mathcal{P})$. Furthermore, the resultant spectral efficiency is formulated by
\begin{equation}
R=\frac{1}{\tau} \Bigg(\left\lfloor\log_2{M\choose K}\right\rfloor + K\log_2{\mathcal{P}}\Bigg) \hspace{2mm}\text{bits/s/Hz}.
\end{equation}

The $u$th frame of our SIM symbols in the frequency domain is represented by
\begin{align}
\Vec{s}_u=[\Vec{s}_{u,0}^T, \Vec{s}_{u,1}^T, ..., \Vec{s}_{u,N-1}^T]^T \in \mathbb{C}^{LM},
\end{align}
where 
\begin{align}
\Vec{s}_{u,l}=[s_{u,l}^{(0)}, s_{u,l}^{(1)}, ..., s_{u,l}^{(M{-}1)}]^T \in \mathbb{C}^{M}\hspace{4mm} (0\leq l<L).
\end{align}
Furthermore, $s_{u,l}^{(m)}$ is the symbol modulated on the $m$th subcarrier in the $l$th group.
The symbol block has to satisfy the energy constraint of $\mathbb{E}[\Vec{s}_u^H\Vec{s}_u]=LM\sigma_s^2$, where $\sigma_s^2$ is the average energy per symbol.
\subsection{EVD-Based Precoding}
\label{sec:system_model:precoding}
In our scheme, the EVD-based precoding technique \cite{osaki-svd-nofdm} is invoked to attain the equivalent ICI-free parallel substreams at the receiver.
Let us define a matrix ${\bf H}\in \mathbb{C}^{LM\times LM}$, which is composed of ICI components, originating due to the non-orthogonality between subcarriers, introduced by reducing the subcarrier-spacing in the proposed scheme.
More specifically, the $k$th-row and $l$th-column entry of ${\bf H}$ is expressed as
\begin{equation}
h_{k,l} = \int_{-\infty}^{\infty}g(t)g^*(-t)e^{j2\pi(l-k)Ft}dt.
\end{equation}
Note that ${\bf H}$ satisfies the positive semi-definite property owing to its Toeplitz structure.
Hence, based on EVD, ${\bf H}$ is factorized to
${\bf H}={\bf Q\Lambda Q}^T$,
where ${\bf \Lambda}=\text{diag}(\lambda_0, \lambda_1, ..., \lambda_{LM-1})\in\mathbb{R}^{LM\times LM}$ is a diagonal matrix containing $LM$ eigenvalues $\lambda_i$ $(0\leq i<LM)$, and ${\bf Q}\in\mathbb{R}^{LM\times LM}$ is an orthonormal matrix containing $LM$ eigenvectors.

At the transmitter, the frequency-domain modulated symbols $\Vec{s}_u$ are precoded by ${\bf QP}$, where ${\bf P}\in\mathbb{R}^{LM\times LM}$ is a real-valued diagonal matrix, called power-allocation (PA) matrix.
I.e., the precoded symbols of the $u$th frame in the frequency domain are expressed as
\begin{align}
\Vec{x}_u &=[x_{u,0}, x_{u,1}, ..., x_{u,N-1}]^T \in\mathbb{C}^{N\times N}\\
&={\bf QP}\Vec{s}_u.
\end{align}
Here, according to~\cite{osaki-svd-nofdm}, the PA matrix ${\bf P}$ is given by
\begin{IEEEeqnarray}{rCL}
\label{PA_matrix}
{\bf P}&=& {\bf \Lambda}^{-\frac{1}{2}} \nonumber\\
&=&\text{diag}\Bigg(\frac{1}{\sqrt{\lambda_0}}, \frac{1}{\sqrt{\lambda_1}}, ..., \frac{1}{\sqrt{\lambda_{N-1}}}\Bigg).
\end{IEEEeqnarray}
As observed from (\ref{PA_matrix}), the eigenvalues $\lambda_i$, $0\leq i<N$ have to be positive and sufficiently large to be tractable~\cite{osaki-svd-nofdm}. Also, low eigenvalues tend to increase the effects of inter-frame interference~\cite{osaki2021wcl}. Hence, the $\tau$ value has to be set sufficiently high to avoid these limitations.

Note that ${\bf H}$ is determined by the system parameters of the subcarrier packing ratio $\tau$ and the subcarrier filters employed at the transmitter. Hence, the EVD can be carried out offline in advance of transmission.

\subsection{Non-Orthogonal FDM Signaling}
\label{sec:system_model:signaling}
Having attained the discrete precoded symbols $\Vec{x}_u$ in the frequency domain, now let us introduce
the transmitted signals in the continuous-time complex-valued baseband representation as follows:
\begin{equation}
\label{nofdm-signal}
x(t)=\sum_{u=-\infty}^{\infty}\sum_{v=0}^{N-1} x_{u,v}g_{u,v}(t),
\end{equation}
where $g_{u,v}(t)=g(t-uT)e^{j2\pi vtF}$ is the basis pulse shape which is obtained by appropriate time-frequency shifting of a prototype transmit filter $g(t)$ in the time-frequency lattice. 
\subsection{Detection Algorithm}
\label{sec:system_model:receiver}
At the receiver, the continuous-time signal at the output of the AWGN channel is expressed as
\begin{equation}
r(t)=x(t)+n(t),
\end{equation}
where $n(t)$ is a complex-valued white Gaussian random process with a zero mean and a spectral density of $N_0$.
Based on the received signal $r(t)$, the receiver carries out matched filtering, sampling, ICI cancellation, and demodulation to recover the transmitted bits.
These steps are described as follows.
\subsubsection{Matched Filtering \& Sampling}
\label{sec:system_model:receiver:matched}
The received signal is projected onto the basis pulse $g_{u,v}$ based on matched filtering to maximize the signal-to-noise ratio (SNR). Then, the matched-filtered signal is sampled. To be more specific, the sampled symbol on the $v$fth subcarrier in the $u$th frame is represented by
\begin{IEEEeqnarray}{rCL}
\label{}
r_{u,v} &=& \sum_{m=-\infty}^{\infty}\sum_{n=0}^{N-1} x_{m,n} \nonumber\\
& & \underbrace{\int_{-\infty}^{\infty} g(t-mT)g^*(t-uT)e^{j2\pi (n-v)tF}dt}_{h_{m,n}} \nonumber\\
&+& \underbrace{\int_{-\infty}^{\infty} n(t)g^*(t-uT)e^{-j2\pi vtF}dt}_{\eta_{u,v}}, \label{eq:sample1}
\end{IEEEeqnarray}
where the third term $\eta_{u,v}$ corresponds to the additive noise component.
From \eqref{eq:sample1}, the vectorial form of the samples symbols are expressed as
\begin{align}
\Vec{r}_u &= [r_{u,0}, r_{u,1}, ..., r_{u,N-1}]^T \in \mathbb{C}^N\\
&={\bf H}\Vec{x}_u + \Vec{\eta}_u, \label{eq:ICIsystem}
\end{align}
where
$\Vec{\eta}_u = [\eta_{u,0},\cdots,\eta_{u,N-1}]^T$.
The noise samples $\eta_{u,v}$ are correlated and have the covariance of 
$\mathbb{E}[\Vec{\eta}_{u}\Vec{\eta}_{u}^H]=N_0{\bf H}$,
which is specific to SEFDM.

\subsubsection{ICI Cancellation}
\label{sec:system_model:receiver:ICI}
Next, ICI-contaminated symbols~\eqref{eq:ICIsystem} are decomposed with the aid of EVD.
By multiplying the orthonormal matrix $\mathbf{Q}^H$ by the received samples $Vec{r}_u$, we obtain
\begin{align}
\label{diagonalization_p1}
\Vec{r}_u^{\dagger} &= \mathbf{Q}^H\Vec{r}_u\\
\label{diagonalization_p2}
&= \mathbf{Q}^H{\bf H}\Vec{x}_u + \mathbf{Q}^H\Vec{\eta}_u\\
\label{diagonalization_p3}
&= \mathbf{Q}^H \underbrace{{\bf Q\Lambda Q}^H}_{\bf H}\underbrace{{\bf Q}\mathbf{P}\Vec{s}_u}_{\Vec{x}_u} + \mathbf{Q}^H\Vec{\eta}_u\\
\label{diagonalization_p4}
&={\bf \Lambda}\mathbf{P}\Vec{s}_u+\mathbf{Q}^H\Vec{\eta}_u. 
\end{align}
Since ${\bf \Lambda}$ and $\mathbf{P}$ are diagonal matrices, the ICI effects are eliminated in \eqref{diagonalization_p4}.
Furthermore, the noise samples $\mathbf{Q}^H\Vec{\eta}$ in \eqref{diagonalization_p4} are uncorrelated with the diagonalized covariance matrix of 
$\mathbb{E}[\mathbf{Q}^H\Vec{\eta\eta}^H\mathbf{Q}]=N_0{\bf \Lambda}$.
As a consequence, the symbols $\Vec{r}_u^{\dagger}$ correspond to $N$ independent substreams with the scaling term of ${\bf \Lambda}\mathbf{P}$ and the uncorrelated additive term of 
$\mathbf{Q}^H\Vec{\eta}$.
Finally, $\Vec{r}_u^{\dagger}$ are equalized with one-tap equalizer as follows
\begin{align}
\Vec{r}_{eq} &= ({\bf \Lambda}\mathbf{P})^{-1}\Vec{r}_u^{\dagger}.\\
&= \Vec{s}_u + ({\bf \Lambda}\mathbf{P})^{-1}\mathbf{Q}^H\Vec{\eta}.
\end{align}
\subsubsection{Demodulation}
\label{sec:receiver:demod}
Next, we introduce two types of demodulators, i.e., the maximum likelihood (ML) detector and the log-likelihood ratio (LLR) based low-complexity detector, for our proposed NOFDM-SIM scheme.

\emph{3.1)} \emph{ML Detection}:
The ML detector attains optimal detection performance by performing an exhaustive joint search on the legitimate SIM and APSK constellation.
The equalized symbols $\Vec{r}_{eq}$ are split into $L$ sub-blocks $\Vec{r}_{eq}^{l}$ $(0\leq l<L)$, each containing $M$ symbols, similar to the transmitter.
Then, the estimated information bits in the $l$th sub-block $\hat{\Vec{b}}^l=(\hat{\Vec{b}}_1^l,\hat{\Vec{b}}_2^l)$ is computed as follows:
\begin{equation}
\label{ml_dec}
\hat{\Vec{b}}^l=\argmin_{(\Vec{b}_1^l,\Vec{b}_2^l)} \left\Vert\Vec{r}_{eq}^{l}-\Vec{s}_u|_{(\Vec{b}_1^l,\Vec{b}_2^l)}\right\Vert^2.
\end{equation}
Naturally, the complexity of ML detection is significantly high, especially when the number of subcarriers $M$ is high.

\emph{3.2)} \emph{Low-Complexity LLR-Based Detection: }
To avoid the high complexity imposed by exhaustive ML search, reduced-complexity two-stage LLR-based detection is used to demodulate the symbols in $\Vec{r}_{eq}$, similar to that employed for OFDM-SIM \cite{basar-ofdm-im}.
More specifically, the LLR of the $m$th symbol $s_m^{(l)}$ in the $l$th frame is given by
\begin{equation}
\label{llr_eq}
L_m^{(l)}=\ln\Bigg(\frac{\sum_{i=1}^P \mathbb{P}(s_m^{(l)}=s_i)|r_m(l)}{\mathbb{P}(s_m^{(l)}=0)|r_m(l)}\Bigg).
\end{equation}
At the first stage of LLR detection, an active index is estimated by a high value of the corresponding LLR. For example, in the case of IM parameters $(M,K,\mathcal{P})=(4,1,4)$, the index with the highest LLR value is detected as the activated index.
The symbol corresponding to the detected active index is then demapped, according to the look-up table.
At the second stage, the APSK symbol of the estimated active symbol is demodulated.

The LLR expression of (\ref{llr_eq}) is further simplified for ease of implementation. Firstly, note that
\begin{IEEEeqnarray}{rCL}
\label{llr_part1}
\sum_{i=1}^P \mathbb{P}(s_m^{(l)}=s_i)&=&\frac{K}{M}\\
\label{llr_part2}
\mathbb{P}(s_m^{(l)}=0)&=&\frac{M-K}{M}.
\end{IEEEeqnarray}
Then, by applying Bayes' theorem to (\ref{llr_eq}) and substituting \eqref{llr_part1}, \eqref{llr_part2} by \eqref{llr_eq}, we arrive at
\begin{IEEEeqnarray}{rCL}
\label{bayes_rule1}
L_m^{(l)} &=& \ln\Bigg(\frac{K}{M-K}\Bigg)+\frac{|r_m^{(l)}|^2}{N_0}\nonumber\\
&+&\ln\Bigg(\sum_{i=1}^P \exp \Big(\underbrace{-\frac{|r_m^{(l)}-s_i|^2} {N_0}}_{z_i}\Big)\Bigg)\nonumber\\
\label{bayes_rule2}
&=&\ln\Bigg(\frac{K}{M-K}\Bigg)+\frac{|r_m^{(l)}|^2}{N_0}
+\ln\Bigg(\sum_{i=1}^P \exp (z_i)\Bigg). \ \ \
\end{IEEEeqnarray}
Furthermore, the last term of (\ref{bayes_rule2}) is simplified with $\text{max}^*()$ function~\cite{viterbi}, which is defined for BPSK modulation as follows:
\begin{align}
\label{max*_func}
\text{max}^*(z_1, z_2)&=\ln(e^{z_1}+e^{z_2})\\
&=\text{max}(z_1,z_2)+\ln\big(1+e^{-|z_1-z_2|}\big).
\end{align}
In general, for $P-$ary APSK modulation, we obtain \cite{viterbi}
\begin{IEEEeqnarray}{rCL}
\label{max*_p-ary}
\text{max}^*(z_1,z_2,...,z_P) 
{=}\text{max}^*(...\text{max}^*(\text{max}^*(z_1,z_2),z_3)...,z_P). \nonumber
\end{IEEEeqnarray}
\section{Performance Results}
\label{performance_analysis}
In this section, we present our performance results to evaluate the PAPR and BER performance of the proposed precoded NOFDM-SIM scheme with and without PA for a channel-uncoded scenario. Two benchmark schemes, i.e., the OFDM with BPSK and the OFDM-SIM with QPSK, were considered. The SIM parameters of $(M,K)=(4,1)$ were employed for the OFDM-IM and the proposed schemes.
A root-raised cosine (RRC) shaping filter was used for OFDM-SIM and the proposed scheme, while a rectangular pulse was employed for OFDM. The number of subcarriers was set to $N=10^3$. For simplicity, we considered a single frame transmission, hence ignoring the effects of inter-frame interference.
\begin{figure}[!t]
\centering
\includegraphics[clip,width=.8\linewidth]{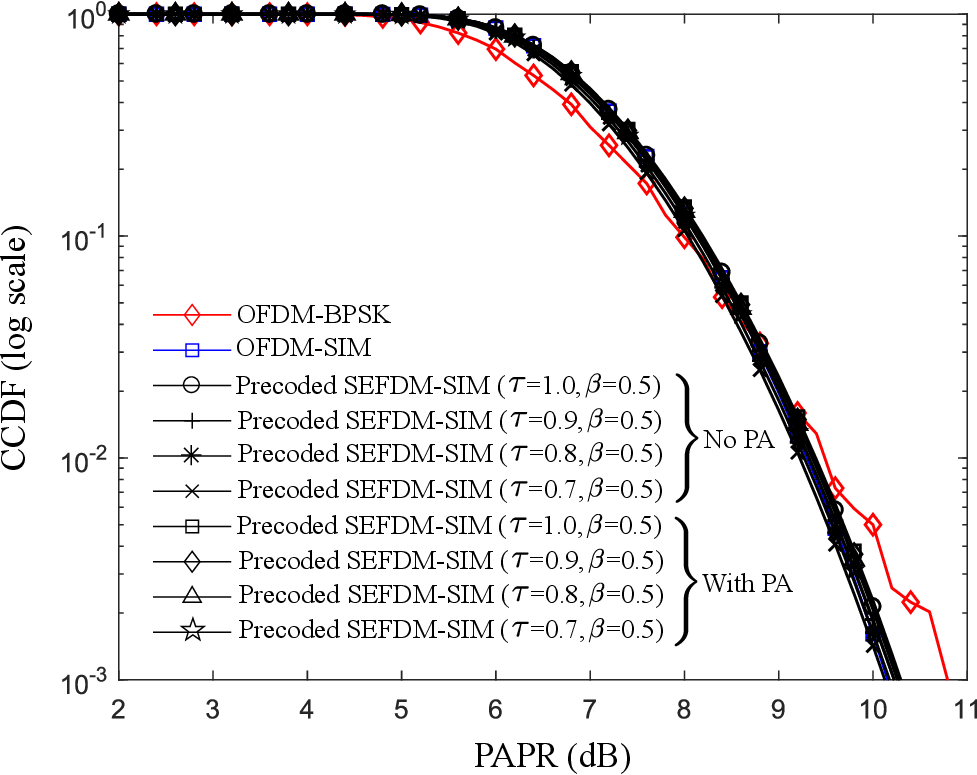}
\caption{CCDF of the PAPR in OFDM, OFDM-SIM and the proposed scheme. The parameters of $(\tau, \beta)=(1,0.5)$, $(0.9, 0.5)$, $(0.8,0.5)$, and $(0.7,0.5)$ were considered for the proposed scheme.}
\label{fig:PAPR_CCDF}
\end{figure}
The PAPR \cite{papr_rahmat}, \cite{papr_han} represents the amplitude fluctuations in the envelope signal of a multicarrier transmission. 
Fig.~\ref{fig:PAPR_CCDF} compares the complementary cumulative distribution functions (CCDF) of the PAPR, where the transmission rate is fixed to 1 bit per channel use. Subcarrier packing ratio $\tau$ and roll-off factor $\beta$ of the RRC shaping filter were set to $(\tau, \beta)=(1,0.5)$, $(0.9, 0.5)$, $(0.8,0.5)$, $(0.7,0.5)$ in order to maintain the positive-definiteness of ${\bf H}$. Observe in Fig.~\ref{fig:PAPR_CCDF} that the proposed scheme exhibited a similar PAPR to other benchmark schemes, regardless of the system parameters employed. Hence, it was found that introducing the non-orthogonality between subcarriers into OFDM-SIM does not increase the PAPR. 

\begin{figure}[!t]
\centering
\includegraphics[clip,width=.78\linewidth]{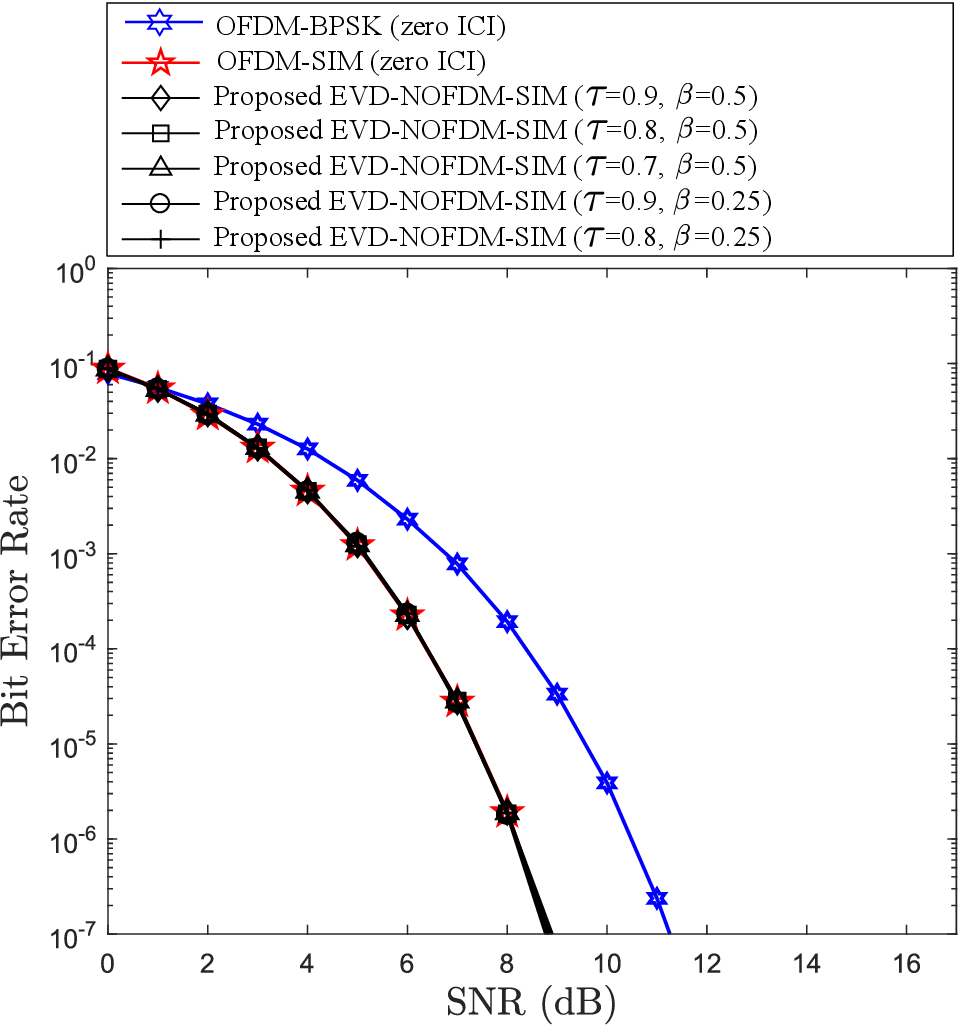}
\caption{BER comparison between OFDM, OFDM-SIM and the proposed scheme with power allocation, where the parameters of $(\tau, \beta)=(0.9, 0.5)$, $(0.8,0.5)$, $(0.7,0.5)$, $(0.9,0.25)$, and $(0.8,0.25)$ were employed for the proposed scheme.}
\label{fig:uncoded_with_PA}
\end{figure}

\begin{figure}[!t]
\centering
\includegraphics[clip,width=.8\linewidth]{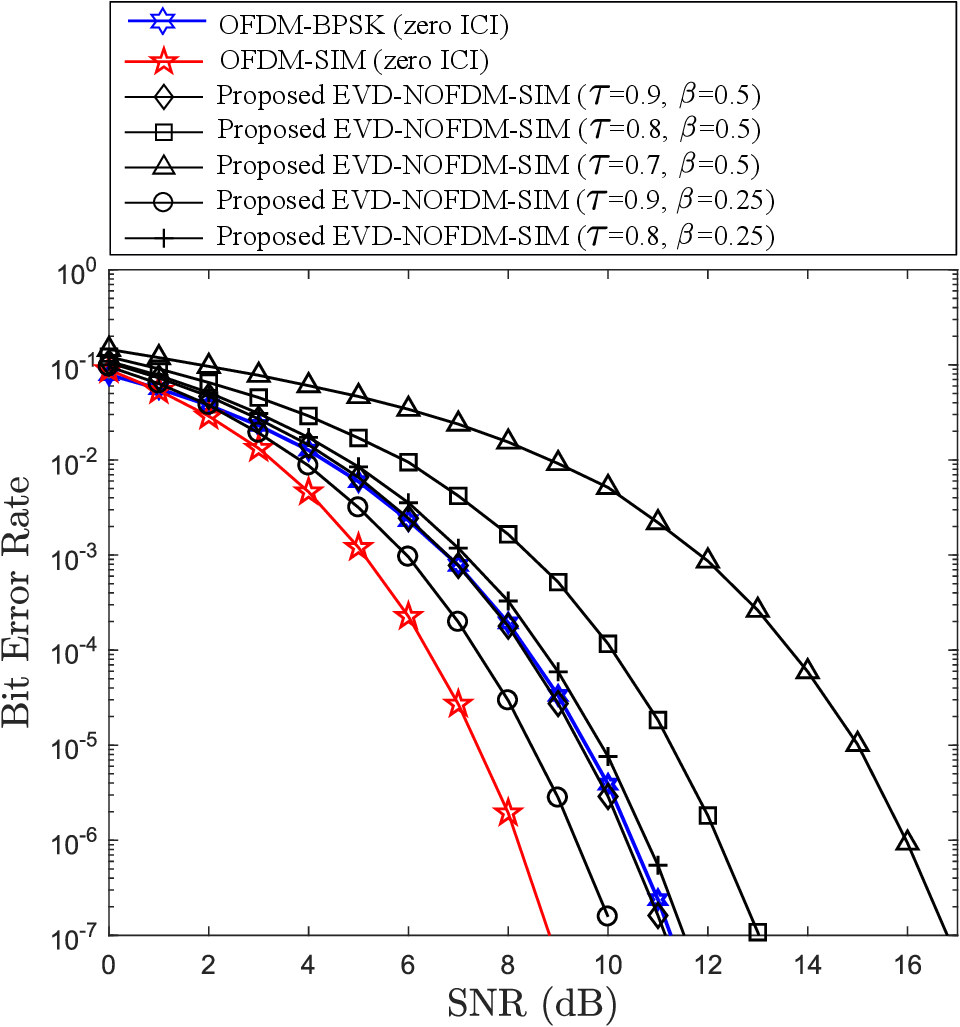}
\caption{BER comparison of OFDM, OFDM-SIM, and the proposed scheme without power allocation, where the parameters of $(\tau, \beta)=(0.9, 0.5)$, $(0.8,0.5)$, $(0.7,0.5)$, $(0.9,0.25)$, and $(0.8,0.25)$ were considered for the proposed scheme.}
\label{fig:uncoded_without_PA}
\end{figure}

Next, in Figs.~\ref{fig:uncoded_with_PA} and \ref{fig:uncoded_without_PA}, we present the BER results of the proposed scheme with and without PA, respectively. ML detection was employed at the receiver of each scheme.
Observed in Fig.~\ref{fig:uncoded_with_PA} that the introduction of PA in the proposed scheme allows us to accomplish perfect cancellation of the ICI effects, hence exhibiting the same BER performance as that of the classic OFDM free from ICI.
This becomes possible owing to the benefits of EVD-based diagonalization as well as PA that enables the same received SNR at each subcarrier.
It was also found in Fig.~\ref{fig:uncoded_without_PA} that the absence of PA imposes the BER performance penalty on the proposed scheme over the OFDM benchmark, which increased upon decreasing $\tau$.
\section{Conclusion}
\label{sec:conclusion}
In this paper, we proposed a novel precoded NOFDM with SIM to harness the combined benefits of NOFDM and SIM to increase spectral efficiency.
With the aid of EVD-based precoding, the detrimental ICI effects induced due to non-orthogonality among subcarriers are removed.
Our numerical results for the channel-uncoded scenario demonstrated that the proposed scheme is capable of achieving the ideal BER performance free from the ICI effects while attaining a higher normalized throughput. 
%
%
\section*{Acknowledgement}
The present study was supported in part by the Japan Science and Technology Agency (JST) Precursory Research for Embryonic Science and Technology (PRESTO) (Grant Number JPMJPR1933).

\bibliographystyle{IEEEtran}
\bibliography{mybibfile}

\end{document}